\documentclass{article}
\usepackage{cite}
\usepackage{graphicx}
\usepackage{dcolumn}

\input{tcilatex}
\begin{document}

\date{}
\title{Gross misinterpretation of a conditionally solvable eigenvalue equation}
\author{Paolo Amore\thanks{%
e--mail: paolo@ucol.mx} \\
Facultad de Ciencias, CUICBAS, Universidad de Colima,\\
Bernal D\'{\i}az del Castillo 340, Colima, Colima,Mexico \\
and \\
Francisco M. Fern\'andez\thanks{%
e--mail: fernande@quimica.unlp.edu.ar} \\
INIFTA, Divisi\'{o}n Qu\'{\i}mica Te\'{o}rica,\\
Blvd. 113 y 64 (S/N), Sucursal 4, Casilla de Correo 16,\\
1900 La Plata, Argentina}
\maketitle

\begin{abstract}
We solve an eigenvalue equation that appears in several papers about a wide
range of physical problems. The Frobenius method leads to a three-term
recurrence relation for the coefficients of the power series that, under
suitable truncation, yields exact analytical eigenvalues and eigenfunctions
for particular values of a model parameter. From these solutions some
researchers have derived a variety of predictions like allowed angular
frequencies, allowed field intensities and the like. We also solve the
eigenvalue equation numerically by means of the variational Rayleigh-Ritz
method and compare the resulting eigenvalues with those provided by the
truncation condition. In this way we prove that those physical predictions
are merely artifacts of the truncation condition.
\end{abstract}

\section{Introduction}

$\label{sec:intro}$

In a series of papers, several authors discussed a wide variety of physical
models in cylindrical coordinates that, after some suitable transformations
of the main dynamical (or eigenvalue) equation, can be reduced to an
eigenvalue equation in the radial variable with Coulomb (or Coulomb--like)
and harmonic (or harmonic-like) interactions. Through further application of
the Frobenius (power-series) method they obtained a three-term recurrence
relation. They stated that in order to have finite solutions (or
normalizable ones) the power series should terminate and they placed a
suitable truncation condition for that purpose. As a result of this
truncation those authors invariably draw the conclusion that some model
parameters, like the intensity of a magnetic field or the oscillator
frequency, for example, should be discrete. In other words, they appeared to
suggest that the eigenvalue equation has square-integrable solutions only
for some particular values (allowed values) of such parameters. Several
years ago, Ver\c{c}in\cite{V91} derived an exact solution to the problem of
two identical charged anyons moving in a plane under the influence of a
static uniform magnetic field perpendicular to that plane. He argued that
there are bound states if and only if the series terminates, which occurs
only for certain discrete values of the magnetic field. Later, Myrheim et al%
\cite{MHV92} discussed Ver\c{c}in's results with more detail finding that
there are solutions for all values of the magnetic-field intensity. Furtado
et al\cite{FDMBB94} discussed the influence of a disclination on the
spectrum of an electron or a hole in a magnetic field in the framework of
the theory of defects and three-dimensional gravity of Katanaev and Volovich%
\cite{KV92}. Although they were aware of the results derived by Myrheim et al%
\cite{MHV92}, they surprisingly concluded that the cyclotron frequency and
the magnetic field should depend on the quantum numbers. From these papers
sprouted many others: for example, Bakke and Moraes\cite{BM12} introduced a
geometric model to explain the origin of the observed shallow levels in
semiconductors threaded by a dislocation density and find allowed values of
the oscillator frequency or of the constant $k$ associated to the momentum
along the $z$-axis. Bakke and Beilish\cite{BB12} obtained the bound states
for a non-relativistic spin-half neutral particle under the influence of a
Coulomb-like potential induced by the Lorentz symmetry breaking effects.
They claimed to present a new possible scenario of studying the Lorentz
symmetry breaking effects on a non-relativistic quantum system defined by a
fixed space-like vector field parallel to the radial direction interacting
with a uniform magnetic field along the $z$-axis. They also discussed the
influence of a Coulomb-like potential induced by Lorentz symmetry violation
effects on a two-dimensional harmonic oscillator and found allowed values of
the cyclotron frequency. Bakke\cite{B14} discussed a model that consists of
the interaction between a moving electric quadrupole moment and a magnetic
field and also added a two-dimensional harmonic-oscillator potential thus
obtaining allowed values of the oscillator frequency. Bakke\cite{B14b}
studied the bound states of a quantum-mechanical model given by the
interaction between the electric quadrupole of a moving particle and an
electric field. In two other models the author added a harmonic potential
and a linear plus harmonic potential and also found allowed oscillator
frequencies. Bakke and Belich\cite{BB14} studied the effects of the Lorentz
symmetry violation in the non-relativistic quantum dynamics of a spin-$\frac{%
1}{2}$ neutral particle interacting with external fields confined to a
two-dimensional quantum ring and also found oscillator frequencies that
depend on the quantum numbers. Fonseca and Bakke\cite{FB15} proposed a model
for the interaction of a magnetic quadrupole moment with electric and
magnetic fields and in a second model they added a harmonic-oscillator
potential finding that the angular frequency depends on the quantum numbers.
Bakke and Furtado\cite{BF15} studied the influence of a Coulomb-type
potential on the Klein-Gordon oscillator and found that the angular
frequency should depend on the quantum numbers. Vit\'{o}ria et al\cite{VBB18}
analyzed the interface between a theory that goes beyond the Standard Model
and quantum mechanics. Based on a line of research that deals with the
violation of the Lorentz symmetry, where it is assumed that at least one
privileged direction in the spacetime, they searched for a hint of a
fundamental theory that goes beyond the energy scale of the Standard Model.
According to the authors, this fundamental theory describes a spontaneous
breaking of the Lorentz symmetry by a tensor background, which can be
considered to be a perturbation. Then, they analyzed the interaction of a
scalar particle with a Coulomb-type potential in the presence of a
background of the violation of the Lorentz symmetry produced by a tensor
field. Vit\'{o}ria and Belich\cite{VB19} investigated the association of the
pointlike global monopole to the point defects in elastic solids were a
harmonic oscillator is immersed. They studied the topology effects of the
medium on the harmonic oscillator and obtained bound states in analytical
form. They also investigated the effects of the Coulomb and linear central
potentials on the harmonic oscillator in an environment with a pointlike
defect and concluded that the angular frequency of the harmonic oscillator
is restricted to discrete values values determined by the quantum numbers of
the system and of the parameter associated to the pointlike topological
defect. Vit\'{o}ria and Belich\cite{VB19} investigated the association of
the pointlike global monopole to the point defects in elastic solids were a
harmonic oscillator is immersed. They studied the topology effects of the
medium on the harmonic oscillator and obtained bound states in analytical
form. They also investigated the effects of the Coulomb and linear central
potentials on the harmonic oscillator in an environment with a pointlike
defect and concluded that the angular frequency of the harmonic oscillator
is restricted to discrete values determined by the quantum numbers of the
system and of the parameter associated to the pointlike topological defect.
Vit\'{o}ria and Belich\cite{VB20b} investigated the relativistic oscillator
model for spin$-1/2$ fermionic fields, in a background of breaking the
Lorentz symmetry governed by a constant vector field inserted in the Dirac
equation by non-minimal coupling and proposed two possible scenarios of
Lorentz symmetry violation which induce a Coulomb type potential. They
determined the relativistic energy profile for the Dirac oscillator
analytically and conjecture an interesting quantum effect: the frequency of
the Dirac oscillator is determined by the quantum numbers of the system and
the parameters that characterize the scenarios of Lorentz symmetry
violation. Vieira and Bakke\cite{VB20} discussed the Aharanov-Bohm effect
for the bound states of a neutral particle with a magnetic quadrupole moment
that interacts with axial fields. They showed that the eigenvalues depend on
the geometric quantum phase, which gives rise to an analog of the
Ahranov-Bohm effect. In addition, they concluded that each radial mode
yields a different set of allowed values of the angular frequency of the
harmonic term included in the interaction of one of the models.

In other applications the radial part of the eigenvalue equation exhibits
Coulomb, linear and harmonic terms. For example, Figueiredo Medeiros and
Becerra de Mello\cite{FB12} analyzed the relativistic quantum motion of
charged spin-0 and spin-$\frac{1}{2}$ particles in the presence of a uniform
magnetic field and scalar potentials in the cosmic string spacetime and
concluded that there must be allowed values of the cyclotron frequency.
Bakke and Belich\cite{BB13b} studied the arising of a Rashba-like coupling,
a Zeeman-like term and a Darwin-like term induced by Lorentz symmetry
breaking effects in the non-relativistic quantum dynamics of a spin-$\frac{1%
}{2}$ neutral particle interacting with external fields arriving at the
conclusion that not all the cyclotron frequencies are allowed. Finally,
Hassanabadi et al\cite{HMM20} studied the interaction of magnetic quadrupole
moment of neutral particle systems (such as atoms or molecules) with a
radial electric field for non-relativistic particles in a rotating frame
that tends to a uniform effective magnetic field perpendicular to the plane
of motion of the neutral particle. To this list we also add some of the
papers cited above\cite{VB19,VB20b}.

The case of linear plus harmonic terms has also been of interest\cite{VB19};
for example, Olivera\cite{OBB20} et al considered the hypothesis of a
privileged direction in the space-time. To this end they take into account a
background of the Lorentz symmetry violation determined by a fixed spacelike
vector field and analyzed quantum effects of this background on the
interaction of a nonrelativistic electron with a nonuniform electric field
produced by a uniform electric charge distribution. They predicted the
existence of allowed angular frequencies.

The purpose of this paper is to study the radial eigenvalue equation derived
in those papers and investigate to which extent the truncation of the power
series by means of the tree-term recurrence relation affects the physical
conclusions drawn by their authors. In section~\ref{sec:time-dependent} we
outline the main equation solved in the papers mentioned above. In section~%
\ref{sec:three-term-rec} we solve the radial eigenvalue equation by means of
the Frobenius method and truncation through a three-term recurrence relation
for the series coefficients in order to derive analytical solutions. By
means of a reliable variational method we also obtain accurate numerical
eigenvalues that are compared with the analytical ones. Finally, in section~%
\ref{sec:conclusions} we summarize the main results and draw conclusions.

\section{The time-dependent equation}

\label{sec:time-dependent}

In several of the papers listed above the starting point is a time-dependent
quantum-mechanical equation of the form\cite
{BM12,BB12,B14,B14b,BB14,FB15,FB12,BB13b,HMM20}
\begin{equation}
i\frac{\partial \psi }{\partial t}=H\psi ,  \label{eq:time_dep_general}
\end{equation}
where the Hamiltonian operator $H=H\left( \rho ,\partial _{\rho },\partial
_{\varphi },\partial _{z}\right) $, $\partial _{q}=\partial /\partial q$, is
given in cylindrical coordinates $0<\rho <\infty $, $0\leq \varphi \leq 2\pi
$, $-\infty <z<\infty $. Upon choosing the particular solution
\begin{equation}
\psi (t,\rho ,\varphi ,z)=e^{-i\mathcal{E}t}e^{ij\varphi }e^{ikz}R(\rho ),
\label{eq:particular_sol_general}
\end{equation}
where $j=l=0\pm 1\pm 2,\ldots $ in some cases\cite{B14,B14b,FB15,BF15}, $%
j=l+1/2$ in others\cite{BM12,BB12,BB14,FB12,BB13b} and $-\infty <k<\infty $
the authors derive an eigenvalue equation for $R(\rho )$:
\begin{equation}
H_{jk}\left( \rho ,\partial _{\rho }\right) R(\rho )=\mathcal{E}R(\rho ).
\label{eq:eigen_R_gen}
\end{equation}

In passing, we mention that in most of the papers listed above the authors
claim to have used units such that $\hbar =c=1$ or the like. This
non-rigorous way of choosing suitable or natural units has been criticized
in a recent pedagogical paper\cite{F20}.

In most of those papers the motion of the particle is unbounded in the $z$
direction and the spectrum is continuous which is plainly reflected in the
dependence of $\mathcal{E}$ on $k$. More precisely, a bound state requires
that
\begin{equation}
\int \int \int \left| \psi (t,\rho ,\varphi ,z)\right| ^{2}\rho \,d\rho
\,d\varphi \,dz<\infty ,  \label{eq:bound-state_def_psi}
\end{equation}
but in the general example outlined above the improper integral over $z$ is
divergent. However, there is interest in the quantization of the motion in
the $x-y$ plane\cite{LL65,CDL77}. On the other hand, other authors consider
a motion in a plane where there are truly bound states\cite{V91,MHV92,BF15}.
In what follows, we focus on the motion of the particle in the $x-y$ plane.

\section{Three-term recurrence relation}

\label{sec:three-term-rec}

In all the papers listed above the authors arrive at an eigenvalue equation
of the form
\begin{eqnarray}
\hat{L}R &=&WR,  \nonumber \\
\hat{L} &\equiv &-\frac{d^{2}}{d\xi ^{2}}-\frac{1}{\xi }\frac{d}{d\xi }+%
\frac{\gamma ^{2}}{\xi ^{2}}-\frac{a}{\xi }+b\xi +\xi ^{2},
\label{eq:eigen_eq_R}
\end{eqnarray}
where $\gamma $, $a$ and $b$ are real numbers and, in general, $\gamma $
depends on the rotational quantum number $l$.

By means of the ansatz
\begin{equation}
R(\xi )=\xi ^{|\gamma |}e^{-\frac{b\xi }{2}-\frac{\xi ^{2}}{2}}P(\xi
),\,P(\xi )=\sum_{j=0}^{\infty }c_{j}\xi ^{j},  \label{eq:R_series}
\end{equation}
we obtain a three-term recurrence relation for the coefficients $c_{j}$:
\begin{eqnarray}
c_{j+2} &=&\frac{b\left( 2\gamma +2j+3\right) -2a}{2\left( j+2\right) \left(
2|\gamma |+j+2\right) }c_{j+1}+\frac{4\left( 2\gamma +2j-W+2\right) -b^{2}}{%
4\left( j+2\right) \left( 2|\gamma |+j+2\right) }c_{j},  \nonumber \\
j &=&-1,0,1,\ldots ,\;c_{-1}=0,\;c_{0}=1.  \label{eq:rec_rel_gen}
\end{eqnarray}

In the papers just mentioned the authors state, in one way or another, that
in order to obtain bound states one has to force the termination conditions
\begin{equation}
W=W^{(n,l)}=2\left( \gamma +n+1\right) -\frac{b^{2}}{4},\,c_{n+1}=0,\,n=1,2,%
\ldots .  \label{eq:trunc_cond}
\end{equation}
Clearly, under such conditions $c_{j}=0$ for all $j>n$ and $P(\xi )$ reduces
to a polynomial of degree $n$. In this way, they obtain analytical
expressions for the eigenvalues $W^{(n,l)}$ and the radial eigenfunctions $%
R^{(n,l)}(\xi )$\cite
{V91,MHV92,FDMBB94,BM12,BB12,B14,B14b,BB14,FB15,BF15,FB12,BB13b,HMM20}. For
the sake of clarity and generality, in this section we use $\gamma $ instead
of $l$ as an effective quantum number because the form of $\gamma $ is not
the same in all those papers.

\begin{figure}[tbp]
\begin{center}
\includegraphics[width=9cm]{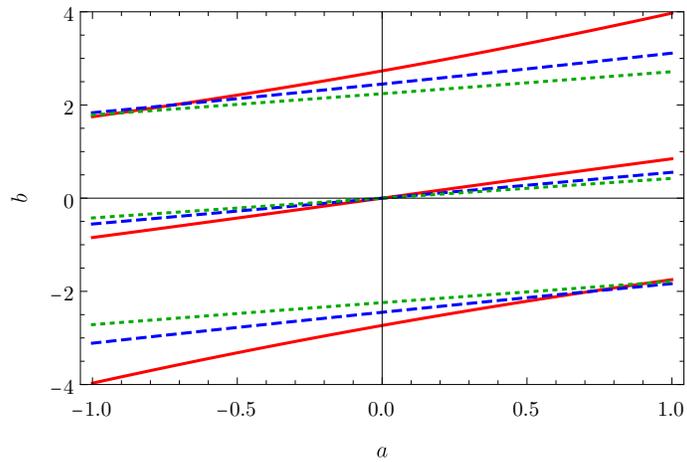}
\end{center}
\caption{Curves $b_{2,\gamma}^{(i)}(a)$, $i=1,2,3$, for $\gamma=0$ (red,
continuous line), $\gamma=1/2$ (blue, dashed line) and $\gamma=1$ (green,
dotted line)}
\label{fig:b(a)}
\end{figure}

For example, when $n=1$ we have
\begin{eqnarray}
W^{(1,\gamma )} &=&2\left( \gamma +2\right) -\frac{b^{2}}{4},\,a_{1,\gamma
}^{(1)}=\frac{2b\left( \gamma +1\right) -\sqrt{b^{2}+8\left( 2\gamma
+1\right) }}{2},  \nonumber \\
a_{1,\gamma }^{(2)} &=&\frac{2b\left( \gamma +1\right) +\sqrt{b^{2}+8\left(
2\gamma +1\right) }}{2},  \label{eq:W,a,n=1}
\end{eqnarray}
or, alternatively,
\begin{eqnarray}
b_{1,\gamma }^{(1)} &=&\frac{2\left[ 2a\left( \gamma +1\right) -\sqrt{%
a^{2}+2\left( 2\gamma +3\right) \left( 2\gamma +1\right) ^{2}}\right] }{%
\left( 2\gamma +1\right) \left( 2\gamma +3\right) },  \nonumber \\
b_{1,\gamma }^{(2)} &=&\frac{2\left[ 2a\left( \gamma +1\right) +\sqrt{%
a^{2}+2\left( 2\gamma +3\right) \left( 2\gamma +1\right) ^{2}}\right] }{%
\left( 2\gamma +1\right) \left( 2\gamma +3\right) }.  \label{eq:b,n=1}
\end{eqnarray}
When $n=2$ we obtain a cubic equation for either $a$ or $b$; for example,

\begin{eqnarray}
W^{(2,\gamma )} &=&2\left( \gamma +3\right) -\frac{b^{2}}{4},  \nonumber \\
&&4a^{3}-6a^{2}b\left( 2\gamma +3\right) +a\left( b^{2}\left( 12\gamma
^{2}+36\gamma +23\right) -16\left( 4\gamma +3\right) \right)  \nonumber \\
&&-\frac{b\left( 2\gamma +1\right) \left( b^{2}\left( 2\gamma +3\right)
\left( 2\gamma +5\right) -16\left( 4\gamma +7\right) \right) }{2}=0,
\label{eq:a,b,n=2}
\end{eqnarray}
from which we obtain either $a_{2,\gamma }(b)$ or $b_{2,\gamma }(a)$; for
example, $a_{2,\gamma }^{(1)}(b)$, $a_{2,\gamma }^{(2)}(b)$, $a_{2,\gamma
}^{(3)}(b)$. In the general case we will have $n+1$ curves of the form $%
a_{n,\gamma }^{(i)}(b)$, $i=1,2,\ldots ,n+1$. Numerical investigation
suggests that all the roots are real; for example, figure~\ref{fig:b(a)}
shows the three curves $b_{2,\gamma }^{(i)}(a)$ for three values of $\gamma $%
.

It is obvious to anybody familiar with conditionally solvable
quantum-mechanical models\cite{D88, BCD17,T16,AF20} (and
references therein) that the approach just described does not
produce all the eigenvalues of the operator $\hat{L}$ for a given
set of values of $\gamma $, $a$ and $b$ but only those states with
polynomial factors $P(\xi )$. It is true that the
approach yields an infinite number of eigenvalues $W^{(n,\gamma )}$, $%
n=0,1,\ldots $, $l=0,\pm 1,\pm 2,\ldots $, but one should understand that $%
W^{(n,\gamma )}$ is an eigenvalue of a set of models given by $a_{n,\gamma
}^{(i)}(b)$, $i=1,2,\ldots ,n+1$whereas $W^{(n^{\prime },\gamma ^{\prime })}$
is an eigenvalue of a set of other models given by the parameters $%
a_{n^{\prime },\gamma ^{\prime }}^{(i^{\prime })}(b)$, $i^{\prime
}=1,2,\ldots ,n^{\prime }+1$. On the other hand, if we solve the eigenvalue
equation (\ref{eq:eigen_eq_R}) in a proper way we obtain an infinite set of
eigenvalues $W_{\nu ,\gamma }(a,b)$, $\nu =0,1,2,\ldots $ for each set of
real values of $a$, $b$ and $\gamma $. The condition that determines these
allowed values of $W$ is that the corresponding radial eigenfunctions $R(\xi
)$ are square integrable
\begin{equation}
\int_{0}^{\infty }\left| R(\xi )\right| ^{2}\xi \,d\xi <\infty ,
\label{eq:bound-state_def_xi}
\end{equation}
Notice that $\nu $ is the actual radial quantum number (that labels the
eigenvalues in increasing order of magnitude) whereas $n$ is just a positive
integer that labels some particular solutions with polynomial factors $P(\xi
)$ (and also a set of quantum mechanical models). In other words: $n$ is a
fictitious quantum number used in those earlier papers as a true one\cite
{V91,MHV92,FDMBB94,BM12,BB12,B14,B14b,BB14,FB15,BF15,FB12,BB13b,HMM20}.

According to the Hellmann-Feynman theorem\cite{CDL77,P68} the actual
eigenvalues $W_{\nu ,\gamma }(a,b)$ of equation (\ref{eq:eigen_eq_R}) are
decreasing functions of $a$ and increasing functions of $b$%
\begin{equation}
\frac{\partial W}{\partial a}=-\left\langle \frac{1}{\xi }\right\rangle ,\,%
\frac{\partial W}{\partial b}=\left\langle \xi \right\rangle .
\label{eq:HFT}
\end{equation}
Therefore, for a given value of $b$ and sufficiently large values of $a$ we
expect negative values of $W$ that the truncation condition fails to
predict. It is not difficult to prove, from straightforward scaling\cite{F20}%
, that
\begin{equation}
W_{\nu ,\gamma }\approx -\frac{a^{2}}{\left( 2\nu +2|\gamma |+1\right) ^{2}}+%
\mathcal{O}\left( a^{-3}\right) ,
\end{equation}
when $a\rightarrow \infty $. What is more, we can conjecture that the pairs $%
\left[ a_{n,\gamma }^{(i)}(b),W^{(n,\gamma )}\right] $, $i=1,2,\ldots ,n+1$
are points on the curves $W_{\nu ,\gamma }(a,b)$, $\nu =0,1,\ldots ,n$,
respectively, for a given value of $b$.

The eigenvalue equation (\ref{eq:eigen_eq_R}) cannot be solved exactly in
the general case. In order to obtain sufficiently accurate eigenvalues of
the operator $\hat{L}$ we resort to the reliable Rayleigh-Ritz variational
method that is well known to yield increasingly accurate upper bounds to all
the eigenvalues of the Schr\"{o}dinger equation\cite{P68} (and references
therein). For simplicity, we choose the basis set of non-orthogonal
functions $\left\{ u_{j}(\xi )=\xi ^{|\gamma |+j}e^{-\frac{\xi ^{2}}{2}%
},\;j=0,1,\ldots \right\} $ and test the accuracy of these results by means
of the powerful Riccati-Pad\'{e} method\cite{FMT89a}.

\begin{figure}[tbp]
\begin{center}
\includegraphics[width=9cm]{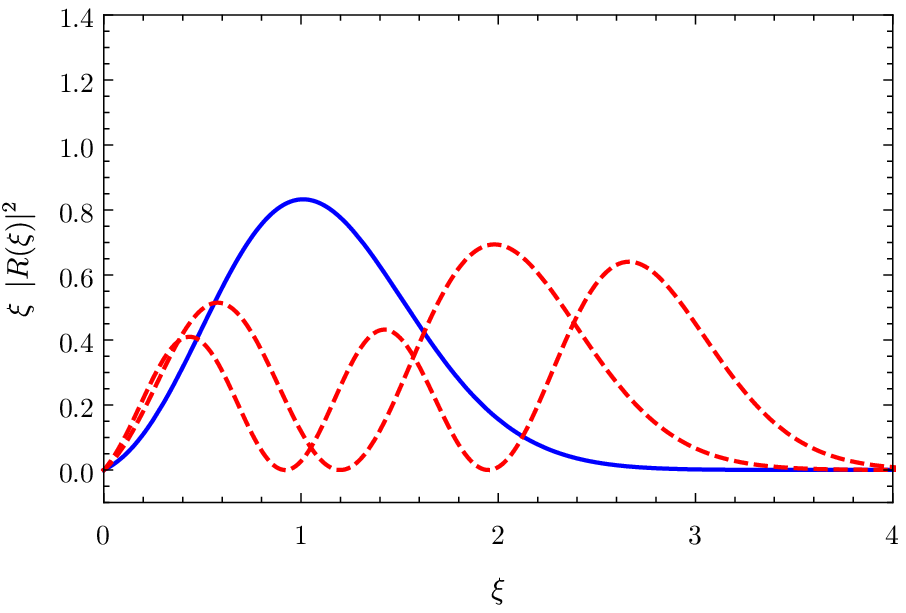}\\[0pt]
\includegraphics[width=9cm]{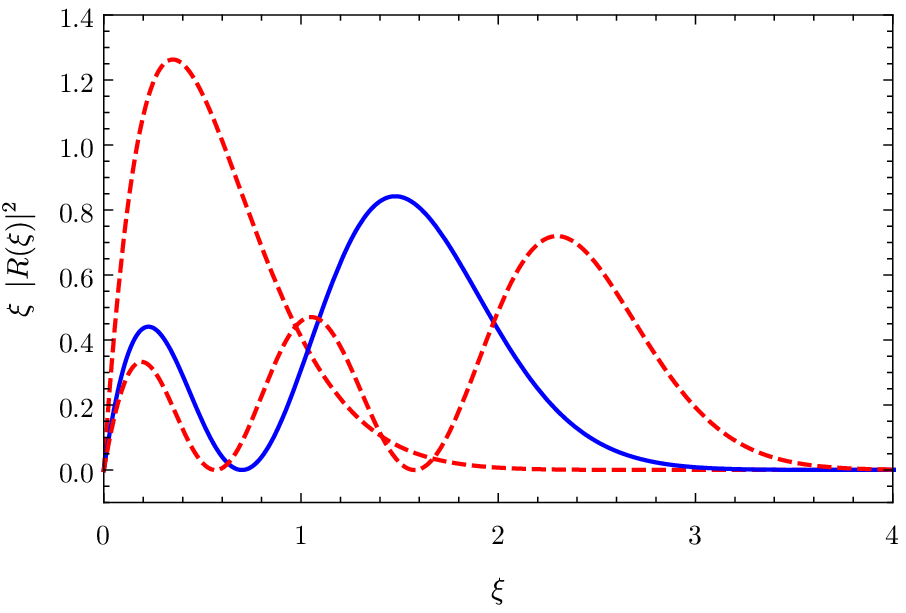}\\[0pt]
\includegraphics[width=9cm]{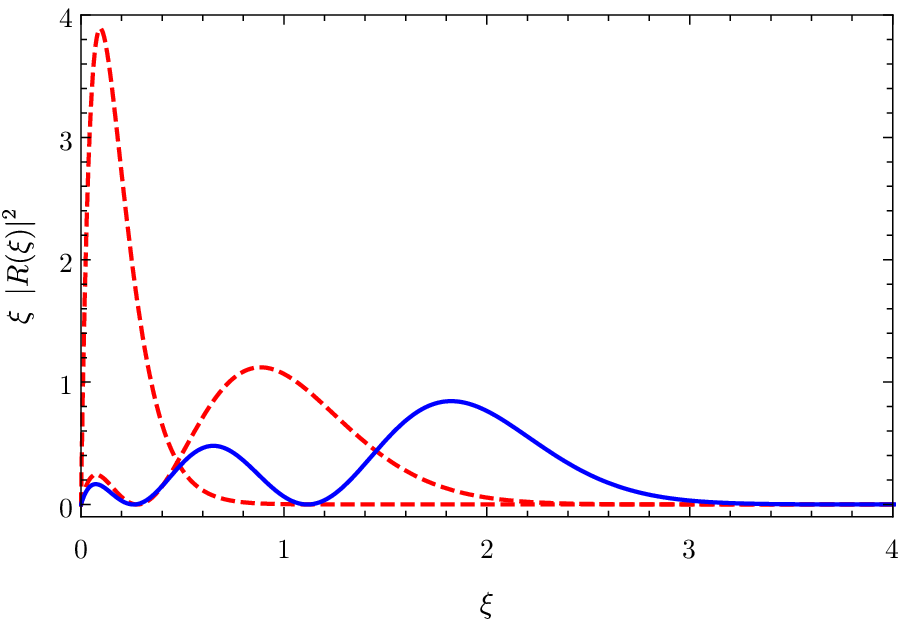}
\end{center}
\caption{Functions $\xi R_{\nu,0}(\xi)^2 $, $\nu=0,1,2$, for $%
a_{2,0}^{(1)}=-1.940551663$ (top), $a_{2,0}^{(2)}=1.190016441$ (center) and $%
a_{2,0}^{(3)}=5.250535221$ (bottom)}
\label{fig:eigenfunctions}
\end{figure}

\begin{figure}[tbp]
\begin{center}
\includegraphics[width=9cm]{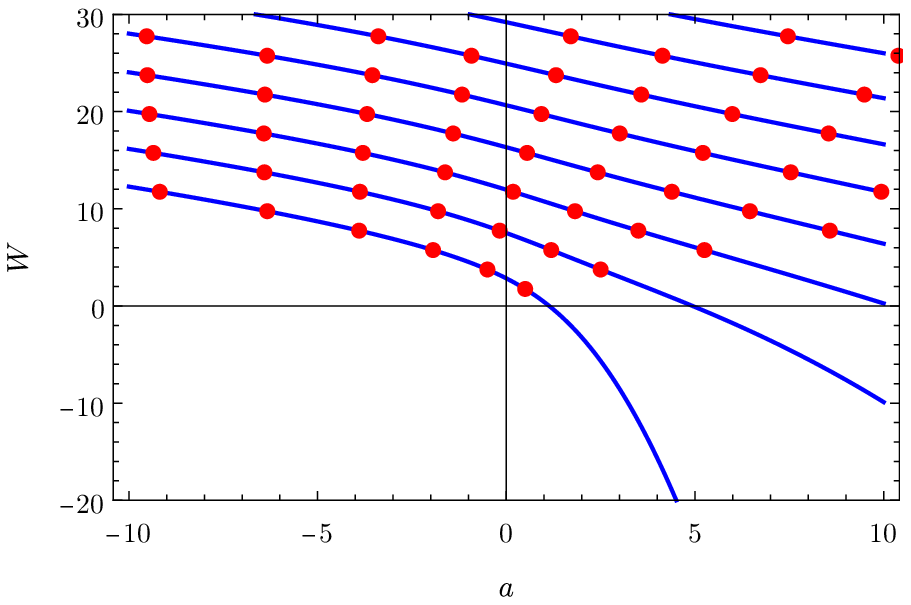}
\end{center}
\caption{Eigenvalues $W_{\nu,0}(a)$ for $b=1$}
\label{fig:W(a)}
\end{figure}

As a first example, we choose $n=2$, $\gamma =0$ and $b=1$ so that $W=5.75$
for the three values of the model parameter $a$: $a_{2,0}^{(i)}$, $i=1,2,3$.
Table~\ref{tab:W(a)} shows the first six eigenvalues for each of these
models. We appreciate that the eigenvalue $W=5.75$ coming from the
truncation condition (\ref{eq:trunc_cond}) is the lowest eigenvalue of the
first model, the second lowest eigenvalue of the second model and the third
lowest eigenvalue for the third model. The truncation condition misses all
the other eigenvalues of each of those models and for this reason it cannot
provide the spectrum of the physical model (given by $a$, $b$, and $\gamma $%
) as believed by the authors of the papers listed above. This table also
shows results for $a=2$ and $b=1$ that is not a point on any curve $a(b)$ \
stemming from the truncation condition. As said above: there are
square-integrable solutions (actual bound states) for any set of real values
of $a$, $b$ and $\gamma $. Figure~\ref{fig:eigenfunctions} shows $\xi R_{\nu
,0}(\xi )^{2}$ for $\nu =0,1,2$ and $a_{2,0}^{(i)}$, $i=1,2,3$. In each
case, the continuous line is the radial function with a polynomial factor $%
P(\xi )$ and the dashed lines indicate the variational results that the
truncation condition does not produce. We appreciate that the truncation
condition provides only one solution for each model and this figure confirms
that the radial eigenfunctions that are missed by the truncation condition
are also square integrable. Figure~\ref{fig:W(a)} shows several eigenvalues $%
W_{\nu ,0}(a)$ for $b=1$. The red points are results given by the truncation
condition and the continuous blue line are variational results that fill the
gaps leaved by such arbitrary condition. These results are most revealing
because they confirm our earlier conjecture about the meaning of the pairs $%
\left[ a_{n,\gamma }^{(i)}(b),W^{(n,\gamma )}(b)\right] $, $n=1,2,\ldots $, $%
i=1,2,\ldots ,n+1$, for a fixed value of $b$. To be clearer: the spectrum of
the operator $\hat{L}$ for $\gamma =0$, $b=1$ and a chosen value of $a$ is
given by the intersections of a vertical line with the blue lines. Such a
vertical line goes through no more than only one red point at best. The
obvious conclusion is that the truncation condition (\ref{eq:trunc_cond})
yields only one eigenvalue of the spectrum of a given model and,
consequently, the allowed cyclotron frequencies, allowed field intensities,
model parameters that depend on the quantum numbers, etc., conjectured by
the authors of the papers listed above, are mere artifacts of this arbitrary
condition. Such claims are nonsensical from a physical point of view.

\section{Conclusions}

\label{sec:conclusions}

In all the papers listed above\cite
{V91,MHV92,FDMBB94,BM12,BB12,B14,B14b,BB14,FB15,BF15, FB12,BB13b,HMM20} the
authors made two basic, conceptual errors. The first one is to believe that
the only possible bound states are those with polynomial factors $P(\xi )$.
We have shown above that there are square-integrable solutions for model
parameters $a$ and $b$ outside the curves $a_{n,\gamma }^{(i)}(b)$ stemming
from the truncation condition (\ref{eq:trunc_cond}). The second one is the
statement that the spectrum of the problem is given by the truncation
condition (\ref{eq:trunc_cond}). It is clear that this equation only
provides one energy eigenvalue for a particular set of models given by the
curves just mentioned. For this reason, all the claims about allowed
cyclotron frequencies, allowed field intensities, model parameters that
depend on the quantum numbers, etc. are nonsensical. All these conclusions
stem from an arbitrary truncation condition that only produces particular
bound states with no physical meaning.

\section*{Acknowledgements}

The research of P.A. was supported by Sistema Nacional de Investigadores
(M\'exico).

\begin{table}[tbp]
\caption{First six eigenvalues $W_{\nu,0}(a)$ for $b=1$ and some values of $%
a $}
\label{tab:W(a)}
\begin{center}
\par
\begin{tabular}{D{.}{.}{1}D{.}{.}{11}}
\hline

\multicolumn{1}{c}{$\nu$}& \multicolumn{1}{c}{$W_{\nu,0}$}  \\
\hline
\multicolumn{2}{c}{$a_{2,0}^{(1)}=-1.940551663$} \\ \hline

0  &    5.75         \\
1  &    9.89404066   \\
2  &    14.06831985  \\
3  &    18.24977457  \\
4  &    22.4306056   \\
5  &    26.60791902  \\  \hline

\multicolumn{2}{c}{$a_{2,0}^{(2)}=1.190016441$} \\ \hline
0  &    -0.1664353619  \\
1  &    5.75           \\
2  &    10.52307155    \\
3  &    15.06421047    \\
4  &    19.4970504     \\
5  &    23.86537389    \\   \hline

\multicolumn{2}{c}{$a=2$} \\ \hline
0  &   -3.230518994    \\
1  &   4.510929109     \\
2  &   9.532275968     \\
3  &   14.1972814      \\
4  &   18.70978427     \\
5  &   23.13559322     \\   \hline

\multicolumn{2}{c}{$a_{2,0}^{(3)}=5.250535221$} \\ \hline
0  &   -27.3245988       \\
1  &   -0.5108147276     \\
2  &   5.75              \\
3  &   10.90599171       \\
4  &   15.71422948       \\
5  &   20.34858964       \\

\hline
\end{tabular}
\par
\end{center}
\end{table}

\begin{thebibliography}{99}
\bibitem{V91}  A. Ver\c{c}in, Phys. Lett. B 260 (1991) 120-124.

\bibitem{MHV92}  J. Myrheim, E. Halvorsen, and A. Ver\c{c}in, Phys. Lett. B
278 (1992) 171-174.

\bibitem{FDMBB94}  C. Furtado, B. G. C da Cunha, F. Moraes, E. R. Bezerra de
Mello, and V. B. Bezzerra, Phys. Lett. A 195 (1994) 90-94.

\bibitem{KV92}  M. O. Katanaev and I. V. Volovich, Ann. Phys. 216 (1992)
1-28.

\bibitem{BM12}  K. Bakke and F. Moraes, Phys. Lett. A 376 (2012) 2838-2841.

\bibitem{BB12}  K. Bakke and H. Belich, Eur. Phys. J. Plus 127 (2012) 102.

\bibitem{B14}  K. Bakke, Ann. Phys. 341 (2014) 86-93.

\bibitem{B14b}  K. Bakke, Int. J. Mod. Phys. A 29 (2014) 1450117.

\bibitem{BB14}  K. Bakke and H. Belich, Eur. Phys. J. Plus 129 (2014) 147.

\bibitem{FB15}  I. C. Fonseca and K. Bakke, J. Math. Phys. 56 (2015) 062107.

\bibitem{BF15}  K. Bakke and C. Furtado, Ann. Phys. 355 (2015) 48-54.

\bibitem{VBB18}  L. L. Vit\'{o}ria, K. Bakke, and H. Belich, Ann. Phys. 399
(2018) 117-123.

\bibitem{VB19}  L. L. Vit\'{o}ria and H. Belich, Phys. Scr. 94 (2019) 125301.

\bibitem{VB20b}  L. L. Vit\'{o}ria, Eur. Phys. J. Plus 135 (2020) 247.

\bibitem{VB20}  S. L. R. Vieira and K. Bakke, Phys. Rev. A 101 (2020) 032102.

\bibitem{FB12}  E. R. Figueiredo Medeiros and E. R. Bezerra de Mello, Eur.
Phys. J. C 72 (2012) 2051.

\bibitem{BB13b}  K. Bakke and H. Belich, Ann. Phys. (Berlin) 526 (2013)
187-194.

\bibitem{HMM20}  H. Hassanabadi, M. de Montigny, and M. Hosseinpour, Ann.
Phys. 412 (2020) 168040.

\bibitem{OBB20}  A. S. Olivera, K. Bakke, and H. Belich, Eur. Phys. J. Plus
135 (2020) 623.

\bibitem{F20}  F. M. Fern\'{a}ndez, Dimensionless equations in
non-relativistic quantum mechanics, arXiv:2005.05377 [quant-ph]

\bibitem{LL65}  L. D. Landau and E. M. Lifshitz, Quantum Mechanics.
Non-relativistic Theory, (Pergamon, New York, 1958).

\bibitem{CDL77}  C. Cohen-Tannoudji, B. Diu, and F. Lalo\"{e}, Quantum
Mechanics, (John Wiley \& Sons, New York, 1977).

\bibitem{D88}  A. DeSousa Dutra, Phys. Lett. A 131 (1988) 319-321.

\bibitem{BCD17}  S. Bera, B. Chakrabarti, and T. K. Das, Phys. Lett. A 381
(2017) 1356-1361.

\bibitem{T16}  A. V. Turbiner, One-dimensional quasi-exactly solvable
Schrodinger equations, arXiv:1603.02992v2

\bibitem{AF20}  P. Amore and F. M. Fern\'{a}ndez, On some conditionally
solvable quantum-mechanical problems, arXiv:2007.03448 [quant-ph]

\bibitem{P68}  F. L. Pilar, Elementary Quantum Chemistry, (McGraw-Hill, New
York, 1968).

\bibitem{FMT89a}  F. M. Fern\'{a}ndez, Q. Ma, and R. H. Tipping, Phys. Rev.
A 39 (1989) 1605-1609.
\end{thebibliography}
\end{document}